\documentclass[preprint,aps]{revtex4-1}
\usepackage{graphicx}
\usepackage{hyperref}
\usepackage{slashed}
\usepackage{rotating}
\usepackage{float}
\usepackage{xcolor}
\usepackage{bm}
\newcommand{\be}{\begin{equation}}
\newcommand{\ee}{\end{equation}}
\newcommand{\bea}{\begin{eqnarray}}
\newcommand{\eea}{\end{eqnarray}}
\newcommand{\ba}{\begin{array}}
	\newcommand{\ea}{\end{array}}

\begin{document}

\title{ Study of vector and  axial-vector form factors and the decay parameters for the semileptonic hyperon decays}
\author{Harleen Dahiya$^{1*}$,  Aarti Girdhar$^2$ and Monika Randhawa$^3$,}
\affiliation{$^1$Department of Physics, Dr. B.R. Ambedkar National
	Institute of Technology, Jalandhar, 144008, India
	\\$^2$ RECAPP,
	Harish-Chandra Research Institute, Chatnaag Road
	Jhunsi, Allahabad 211019, India
\\$^3$ University Institute of Engineering and Technology,
Panjab University, Chandigarh, 160014, India}

\begin{abstract}
Using the standard parametrization of the dipole form,  we have studied the vulnerability of $Q^2$ on the vector form factors ($f_i^{B_iB_f}(Q^2)$) and axial-vector form factors ($g_i^{B_iB_f}(Q^2)$), $i=1,2,3$ computed for the  semileptonic  $B_i \rightarrow B_f l \bar{\nu}$ decays for hyperons	in the framework of chiral constituent quark model ($\chi$CQM).  Both, strangeness changing as well as strangeness conserving decays have been examined.  We also present the dependence of the ratio of hyperon semileptonic decay constants $g_1(Q^2)/f_1(Q^2)$ for these decays. Further, we calculate the CKM matrix elements $V_{ud}$ from strangeness conserving and $V_{us}$ from strangeness changing hyperon decays.  
\end{abstract}

\maketitle
%\pagebreak

\section{Introduction}
Hyperon semileptonic decays not only provide a profound understanding with regard to the internal structure of the hadrons, but are also important for investigating the detailed  dynamics of the hyperons. They are important to recognize the fascinating exchange connecting the weak interactions on one side and the strong interactions on the other side \cite{Gaillard:1984ny,Cabibbo:2003cu,Cabibbo:2003ea} therefore providing an aid to assess New Physics. The hyperon semileptonic decays act as the only available source for obtaining the axial-vector coupling parameters and can be used to test the Standard Model (SM). They play a deciding role in determining $V - A$ structure and provide independent constraints and also open a possibility of estimating the Cabibbo Kobayashi Maskawa (CKM) quark mixing matrix. Owing to the short lifetimes of the hyperons, it is difficult to measure their properties from experiments.  Even though not much data is available on the measurements of form factors, branching fractions, angular interactions and rates of various decays, they provide a pivotal medium to examine the functioning of strong interactions and their degrees of freedom at low energies.

Both vector ($f_{i=1,2,3}$) and axial-vector ($g_{i=1,2,3}$) form factors support details regarding the structural and formalistic hadron constituents. Ever since the measurements of the polarized structure functions of proton $g_1^p(x)$,  the ratio of axial-vector to vector coupling constants of the neutron $\beta-$decay has been determined precisely at zero-momentum transfer.  The baryonic  axial-vector form factors for weak decays have received much attention since  they are most general  to examine the effects arising because of the spontaneous breaking of chiral symmetry.  These correlate the deep inelastic scattering (DIS) data to the quark spin effects.  Using vector and axial-vector form factors, we wish to calculate the two elements of the first row of the $3 \times 3 $ quark mixing CKM matrix, $V_{ud}$ and $V_{us}$, connected by the unitarity condition $|V_{ud}|^2+|V_{us}|^2+|V_{ub}|^2=1$.
The precise measurement of the elements of the CKM matrix elements is crucial as they are the elementary variables of the SM. The super allowed pure Fermi Beta Decay experiment provide $V_{ud}$ and, $V_{us}$ is obtained through semileptonic decays of kaons \cite{Hardy:2014qxa,PDG}. Another way of getting both $V_{ud}$ and $V_{us}$ experimentally is from hyperon semileptonic decays. However, in this case we require additional information of both vector, $f_{i=1,2,3}$ and axial-vector, $g_{i=1,2,3}$ form factors.  It is pivotal to get the accurate measurement of $V_{ud}$ and $V_{us}$ since the precision test of the first row of the CKM matrix elements depends on the uncertainty in the values of $|V_{ud}|$ and $|V_{us}|$. $|V_{ub}|$  being very small ($|V_{ub}|=(3.82\pm 0.24)\times 10^{-3}$ \cite{PDG}) has very little impact on the unitarity condition.

%%%%%%%%%%%%%%%%%%%%%%%%%%%%%%%%%%%%%%%%%%%%%%%%%%%%%%%%%%%%%%%%%%

 Although SU(3) flavor symmetry was presumed while probing the data of vector and axial-vector coupling constants in the initial experiments \cite{cern-WA2}, the eventually performed experiments  distinctly exhibited broken SU(3) symmetry which was initially reported from the estimation corresponding to the $\Sigma^- \rightarrow n e^{-} \bar{\nu_e}$ decay extracting $ |(g_1(0) - 0.133g_2(0))/f_1(0)|$ = 0.327 $\pm 0.007 \pm 0.019$   giving
$\frac{g_1(0)}{f_1(0)}=-0.20 \pm 0.08$ and $\frac{g_2(0)}{f_1(0)}=-0.56 \pm
0.37$ \cite{syh}. The distinction with the SU(3) symmetric results was quite substantial ($\frac{g_1(0)}{f_1(0)}=-0.328
\pm 0.019$ and $\frac{g_2(0)}{f_1(0)}=0$). The most recent experimental data accessible for $\Sigma^- \rightarrow n e^{-}
\bar{\nu_e}$ decay is  $ |(g_1(0) -
0.237g_2(0))/f_1(0)|$ = 0.340 $\pm 0.017$ and   $\frac{f_2(0)}{f_1(0)}=0.97 \pm
0.14$ \cite{PDG}. 
%%%%%%%%%%%%%%%%%%%%%%%%%%%%%%%%%%%%%%%%%%%%%%%%%%%%%%%%%%%%%%%%%%%%%%%%%%%%%%%%
For the decay $\Xi^0 \rightarrow \Sigma^+ e^-
\bar {\nu_e}$,  form factors estimations were published by Fermilab KTeV experiment  (E799) \cite{ktev}. They presented $\frac{f_2(0)}{f_1(0)}$=
2.0$ \pm 1.2_{stat} \pm 0.5_{syst}$ and $\frac{g_1(0)}{f_1(0)}$=
1.32$^{+0.21}_{-0.17stat} \pm 0.05_{syst}$ assuming SU(3) symmetry. The NA48/1 Collaboration published the data on $\Xi^0 \rightarrow \Sigma^+ e^-\bar {\nu_e}$ decay \cite{batley} with somewhat improved statistics, for example, $\frac{g_1(0)}{f_1(0)}$ =
1.20 $\pm$ 0.05. For the same decay, the  PDG results \cite{PDG} with SU(3) symmetry breaking presumption are $\frac{g_1(0)}{f_1(0)}$ =
1.22 $\pm$ 0.05,  $\frac{g_2(0)}{f_1(0)}$=
$-1.7^{+2.1}_{-2.0stat} \pm 0.5_{syst}$ and $\frac{f_2(0)}{f_1(0)}$=
2.0$ \pm 0.9$.  For the other member of the isospin pair, the decay  $\Xi^- \rightarrow \Sigma^0 e^-
\bar {\nu_e}$ measured $\frac{g_1(0)}{f_1(0)}= 1.25^{+0.14}_{-0.16stat} \pm 0.05_{syst}$ \cite{cascade-sigma}. The results of $\frac{g_1(0)}{f_1(0)}$ for $\Lambda$ and  $\Xi^-$ decays are respectively  $\Lambda \rightarrow p e^- \bar {\nu_e}=-0.718 \pm 0.015$ and  $\Xi^- \rightarrow \Lambda e^-\bar {\nu_e}=-0.25 \pm 0.05$ \cite{PDG}. We also have  $\frac{f_2(0)}{f_1(0)}$=
0.01$ \pm 0.10$  \cite{PDG} from the 
$\Sigma^- \rightarrow \Lambda e^-\bar {\nu_e}$ and $\Lambda \rightarrow p e^- \bar {\nu_e}$ decays \cite{Lambda-p,cascade-Lambda}. Future experiments are planning to measure the data on hyperons, for example, at CERN in the NA62 experiment  \cite{NA62-furture}, at PANDA in the $p\bar{p}$ collider  \cite{PANDA} and at FAIR/GSI \cite{J-PARC}.
%%%%%%%%%%%%%%%%%%%%%%%%%%%%%%%%%%%%%%%%%%%%%%%%%%%%%%%%%%%%%%
Theoretically, various studies have been carried out to understand SU(3) symmetry breaking. For example, chiral
perturbation theory (ChPT) \cite{lac,ruben1}, lattice QCD \cite{gaud}, $1/N_c$ expansion of QCD  \cite{Flores-Mendieta:2004cyh,Mateu:2005wi} but in spite of the progress, there is still a need for more clarity and improvement.
$V_{us}$ has been calculated in the framework of Cabibbo Model in \cite{Cabibbo:2003cu,Cabibbo:2003ea} for SU(3) symmetric case and in the relativistic constituent quark model by taking in account SU(3) breaking \cite{Schlumpf:1994fb,Garcia:1991pu}. The calculation of $V_{us}$ has been carried out with SU(3) breaking effects from the hyperon decays in \cite{SDC}. Computations in the $1 \over N_{c}$ expansion of QCD have been done by taking in account the SU(3) symmetry breaking  \cite{Flores-Mendieta:2004cyh,Mateu:2005wi} and mass splitting interactions \cite{Yamanishi:2007zza}.
%%%%%%%%%%%%%%%%%%%%%%%%%%%%%%%%%%%%%%%%%%%%%%%%%%%%%%%%

A number of theoretical and experimental techniques have evolved, since the earlier chiral constituent quark model ($\chi$CQM) estimations for hyperon decay parameters   \cite{manohar,cheng,johan,song,hd,nsweak,SDC} where the dynamics of light quarks is explained through an effective Lagrangian approach. The $\nu \bar{\nu}$ elastic  scattering \cite{antineutrino1,antineutrino2}, the electro-production of pion on the proton \cite{pion-electro} and  the Miner$\nu$a experiment \cite{minerva} made an effort to interpret the vulnerability of $Q^2$ on the form factors corresponding to vectors and axial-vectors.  All these efforts lead to  more precision in data. Taking clue from the above developments, the vulnerability of $Q^2$  explored for the $f_i^{B_iB_f}(Q^2)$ (vector) and  $g_i^{B_iB_f}(Q^2)$ (axial-vector) form factors would be worth the effort. This interpretation  would undoubtedly furnish a benchmark to test the key features of the model in the nonperturbative regime of QCD. This can be further applied to deduce the Cabibbo–Kobayashi–Maskawa (CKM) matrix element $V_{us}$ and $V_{ud}$ for the case of both strangeness changing and strangeness conserving hyperon semileptonic decays calculated at $Q^{2}=0$.

In the current communication, we aim to make use of the $\chi$CQM estimations for the 
vector form factors $f_{i=1,2,3}$  and axial-vector form factors $g_{i=1,2,3}$   corresponding to the hyperon semileptonic decays $B_i \rightarrow B_f l \bar{\nu}$ and further calculate the CKM matrix element $V_{ud}$ and $V_{us}$.  The semileptonic decays of the hyperons considered here are both strangeness changing as well as strangeness conserving. Using these form factors, the  standard parametrization of the dipole form will be suitably applied to explain the vulnerability of $Q^2$ on the form factors for vectors ($f_i^{B_iB_f}(Q^2)$) and form factors for axial-vectors ($g_i^{B_iB_f}(Q^2)$) as well as their decay constants.

This article is organized as follows. Section \ref{s2} contains the basics of the framework used and the methdology adopted for the current work. Section \ref{s3} contains the input variables. Section \ref{s4} contains the results and discussion followed by section \ref{s5} providing the summary and outlook of the work.
%%%%%%%%%%%%%%%%%%%%%%%%%%%%%%%%%%%%%%%%%%%%%%%%%%%%%%%%%%%%%%%%%%%
\section{Methodology}\label{s2}
%   {The vector and axial-vector form factors in the chiral constituent quark model}
Taking motivation from the successes of $\chi$CQM model in context of ``proton spin crisis'', magnetic moments of light and charmed baryons, distribution functions of quarks etc. \cite{hd,hdmagnetic}, we calculate the first row CKM matrix elements in this framework, $V_{us}$ from strangeness changing hyperon decays and $V_{ud}$ from strangeness conserving decays.  We also explore the vulnerability of $Q^2$ on the form factors corresponding to vector and axial-vector currents computed in the $\chi$CQM \cite{nsweak}. 

Taking  $B_{i}$ and $B_{f}$ as the baryon initial and final states respectively, the  semileptonic decays can be expressed as \be
B_i \rightarrow B_f l \overline {\rm \nu}_{l},
\ee where $l$ is the lepton ($e$, $\mu$ or $\tau$) and $\overline {\rm \nu}_{l}$ is the complementary antineutrino. In the current work, we have taken $l$ and $\overline {\rm \nu}_{l}$ as  $e$ and $\overline {\rm \nu}_{e}$ respectively. The matrix element corresponding to the process is expressed as \cite{Carson:1987gb},\cite{Ohlsson:1999tr}  
\begin{equation}
	M={G^{2}_{F} \over \sqrt{2}}V_{{q_{i}}{q_f}}\langle B_{f}(p_{f})|J^{\mu}_{h}|B_{i}(p_{i})\rangle \times ({\overline{\rm u}}_{e}(p_{e})\gamma_{\mu}(1-\gamma_{5})u_{\nu_e}(p_{\nu_{e}})),\label{matrixe}
\end{equation}
where $u_{\bar{\nu_e}}(p_{\nu_e})$ and $\bar{u}_e(p_e)$ are the standard neutrino and electron Dirac spinors respectively which are defined in the momentum space, $G_F$ is the Fermi coupling constant, $V_{q_iq_f}$ is the CKM element,  $V_{ud}$ for the decay where strangeness is conserved ($\Delta S=0$) and $V_{us}$ for the decay where strangeness changes ($\Delta S=1$). 
The weak hadronic current, $J^{\mu}_{h}$ can further divided into explicit vector and axial-vector currents  as $J^{\mu}_{h}= J^{\mu}_{V}-J^{\mu}_{A}$. $V^{\mu,a}=\overline{{\bf q}}\gamma^\mu \frac{\lambda^a}{2} {\bf q}$ and  $A^{\mu,a}=\overline{{\bf q}}\gamma^\mu \gamma_5\frac{\lambda^a}{2} {\bf q}$ are the vector and axial-vector currents respectively and 
\bea
\langle B_f|J^{\mu}_h|B_i \rangle&=&\langle B_f|V^{\mu,a}|B_i \rangle -\langle B_f|A^{\mu,a}|B_i \rangle \nonumber \\
&=& \langle B_f|\overline{{\bf q}}\gamma^\mu \frac{\lambda^a}{2} {\bf q}|B_i \rangle-\langle B_f|\overline{{\bf q}}\gamma^\mu \gamma_5\frac{\lambda^a}{2} {\bf q}|B_i \rangle.
\eea
The matrices of the group SU(3) defined by Gell-Mann, relate to the light quark flavor structure and are represented as $\lambda^a$ in the above equation. For strangeness conserving  decays ($\Delta S=0$ transitions), the index $a$ takes the value $1 \pm i 2$ and for strangeness changing decays ($\Delta S=1$ transitions), the index $a$ takes the value $4 \pm i 5$.

With regard to the  vector functions $f_i^{B_iB_f}(Q^2)$ ($i=1,2,3$), the matrix element for vector current can be illustrated through the matrix elements \cite{tommy,Renton:1990td} \be \langle B_f|\overline{{\bf q}}\gamma^\mu \frac{\lambda^a}{2} {\bf q}|B_i \rangle=\bar u_f(p_f) \left( f_1^{B_iB_f}(Q^2)
\gamma^\mu + \frac{f_2^{B_iB_f}(Q^2)i \sigma^{\mu\nu} q_\nu}{M_{B_i}+M_{B_f}} 
+\frac{f_3^{B_iB_f}(Q^2)q^\mu}{M_{B_i}+M_{B_f}} \right) \frac{\lambda^a}{2} u_i(p_i)\,, \label{jv} \ee
whereas with regard to the axial-vector functions $g_i^{B_iB_f}(Q^2)$ ($i=1,2,3$), the matrix element for the axial-vector current  is expressed as \cite{tommy,Renton:1990td}
\be
\langle B_f|\overline{{\bf q}}\gamma^\mu \gamma_5\frac{\lambda^a}{2} {\bf q}|B_i \rangle= \bar u_f(p_f)\left
( g_1^{B_iB_f}(Q^2)\gamma ^\mu  +\frac{g_2^{B_iB_f}(Q^2)i\sigma^{\mu\nu} q_\nu}{M_{B_i}+M_{B_f}}  +\frac{g_3^{B_iB_f}(Q^2)q^\mu}{M_{B_i}+M_{B_f}}  \right) \gamma^5 \frac{\lambda^a}{2} u_i(p_i)\,.\label{ja} \ee 
The transfer of four momenta  can be expressed as $Q^2 = -q^2$, where $ q \equiv
p_i - p_f$.  $M_{B_i}$ $(M_{B_f})$ are the initial (final) baryon states masses. The functions $f_i^{B_iB_f}(Q^2)$ and $g_i^{B_iB_f}(Q^2)$ ($i=1,2,3$)
correspond to the vector and axial-vector form factors which are necessarily real by $G-$parity. To be more explicit, $f_1$ is defined as the form factor arising because of the vector current, $f_2$ (also referred as weak magnetism) arises from the induced tensor current, $f_3$ arises from the induced scalar current, $g_1$ arises from the axial-vector current, $g_2$ (also referred as weak electricity) arises from the induced pseudotensor current  and
$g_3$ arises from the induced pseudoscalar scalar current. Under $G$-parity transformation, opposite sign is obtained for the case of $f_3$ and $g_2$   \cite{wein}.
It is important to mention here that $f_1$ and $g_1$ give the respective coupling constants using the Clebsch-Gordan coefficients through the Ademollo-Gatto theorem for a vanishing four momentum transfer. Since the magnitude  $f_3$ and $g_2$ is very small because of reversal under $G$-parity we can ignore them in the present case.

Another type of  form factors called the Sachs-type, connected to the time and space elements, can be introduced to work out the vector and axial-vector functions at $Q^2=0$ GeV$^2$ \cite{tommy,larry}. The Sachs form factor used in vector functions are $G_{V,0}^{B_iB_f}$ which can be  computed using the time component $\langle B_f|V^{0,a}|B_i \rangle $), whereas $G_{V,V}^{B_iB_f}$ and $G_{V,A}^{B_iB_f}$ can be computed using the space component $\langle B_f|V^{j,a}|B_i \rangle $). For the axial-vector functions, $G_{A,0}^{B_iB_f}$, $G_{A,S}^{B_iB_f}$ and $G_{A,T}^{B_iB_f}$ can be computed using the time $\langle B_f|A^{0,a}|B_i \rangle $ and space components $\langle B_f|A^{j,a}|B_i \rangle $.

To begin with, the light quark  ($u$, $d$, and $s$) dynamics can perhaps be illustrated through the Lagrangian of QCD. At the scale of around 1 GeV, spontaneous breaking of chiral symmetry takes place. As a result of this symmetry breaking a set of massless particles  comes in existence. These particles are identified as $\pi$, $K$, $\eta$ mesons and are referred to as  Goldstone bosons (GBs). If we consider just the effective part of the interaction, the Lagrangian in that case can then  be expressed as \be {\cal
	L}_{{\rm int}} = c_8 { \bar \psi} \Phi {\psi} + c_1{ \bar \psi}
\frac{\eta'}{\sqrt 3}{\psi}= c_8 {\bar \psi}\left( \Phi + \zeta
\frac{\eta'}{\sqrt 3}I \right) {\psi }=c_8 {\bar \psi} \left(\Phi'
\right) {\psi} \,, \label{lagrang4} \ee where $I$ is the usual $3\times 3$
unit matrix and the coupling strength ratio for the $\eta'$ singlet and octet of GBs is $\zeta=c_1/c_8$, $c_1$.

The features of $\chi$CQM have already been presented in Ref. \cite{hd}, however for the sake of continuity  we present here very briefly some essential details. The process describing fluctuation leading to the operative Lagrangian in the $\chi$CQM  \cite{manohar} is
\be q^{\pm} \rightarrow {\rm GB} + q^{'
	\mp} \rightarrow (q \bar q^{'}) +q^{'\mp}\,. \label{basic}
\ee
The basic process, in the $\chi$CQM, is the emission of a GB from the constituent quark. This GB further splits into a $q \bar q^{'}$ pair giving  $q \bar q^{'} +q^{'}$ after the fluctuation. These  $q \bar q^{'} +q^{'}$ quarks (after the fluctuations) constitute the overall sea of quarks created \cite{cheng,johan,hd}. The GB field could be illustrated in connection with the GBs and their probabilities of transition which are suggested by taking nondegenerate quark masses $M_s > M_{u,d}$, nondegenerate GB masses $M_{K},M_{\eta}> M_{\pi}$ and  $M_{\eta^{'}} > M_{K},M_{\eta}$. These variables represent quantitatively the scale to which the sea quarks contribute to the quark
structure of the baryons.

In the presence of a pseudoscalar field $\Phi'$, the  vector current (to lowest order) in $\chi$CQM for the
quark sector is given as
\be
J^\mu_{V,qq'} = \bar q'\gamma^{\mu} q\,, \ee and the axial-vector
current is given as
\be
J^\mu_{A,qq'} =g_a \bar q'\gamma^{ \mu} \gamma^5q- f_{\Phi'}
\partial^{\mu} \Phi'\,, \label{a1} \ee
where $g_a=1$ is the quark axial-vector current coupling constant \cite{wein}, $f_{\Phi'} $ represents
the pseudoscalar decay constant.

In the nonrelativistic limit, the current operators of the three quarks are added for the case of baryons. In this case, we can make use of the quark's Sachs form
factors  to derive the
corresponding Sachs form factors for the baryons. We consider the following strangeness changing hyperon decays: $\Sigma^{-} \rightarrow ne^{-}\bar{\nu}_{e} $, $\Xi^{-}  \rightarrow \Sigma^{0} e^{-}\bar{\nu}_{e}$, $\Lambda  \rightarrow pe^{-}\bar{\nu}_{e} $, $\Xi^{-}  \rightarrow \Lambda e^{-}\bar{\nu}_{e}$,  $\Xi^{0}  \rightarrow \Sigma^{+}e^{-}\bar{\nu}_{e}$  and $n \rightarrow pe^{-}\bar{\nu}_{e}$, 
	$\Sigma^{-} \rightarrow \Sigma^{0} e^{-}\bar{\nu}_{e}$,
	$\Sigma^{-} \rightarrow \Lambda e^{-}\bar{\nu}_{e} $,
	$\Sigma^{+} \rightarrow \Lambda e^{-}\bar{\nu}_{e}$,
	$\Xi^{-}  \rightarrow \Xi^{0} e^{-}\bar{\nu}_{e}$ corresponding to strangeness conserving decay. For details of the form factor calculations, we refer the readers to Ref. \cite{nsweak} and the references therein.

Further, making use of the transition amplitude from Eq. (\ref{matrixe}) and the total decay rate, the matrix element $V_{{q_{i}}{q_f}}$ can be calculated  \cite{Garcia:1985xz} and is given as
\bea
R&=&G^{2}_{F}{\Delta{M}^{5}|V_{us}|^{2} \over 60{\pi}^{3}}\bigg\{ \bigg(1-{{3\over2}E}+{{6\over7}E^{2}}\bigg)f^{2}_{1}+{{4\over7}E^{2}f^{2}_{2}} \nonumber\\
&&+\bigg(3-{9\over2}E+{12\over7}E^{2}\bigg)g^{2}_{1}
+{12\over7}E^{2}g^{2}_{2}+{6\over7}E^{2}f_{1}f_{2}\nonumber\\
&&+\bigg(-4E+6E^{2}\bigg)g_{1}g_{2}\bigg\},\label{Decayrate}
\eea
where $E = {\Delta M \over \Sigma M }$, $ \Delta M = M_{i}-M_{f}$ and $\Sigma M = M_{i}+M_{f}$. 
As mentioned before, for the case where strangeness changes with $\Delta S=1$,  $V_{{q_{i}}{q_f}}$ corresponds to $V_{us}$ and for the case where strangeness is conserved with $\Delta S=0$ it corresponds to $V_{ud}$. In the above equation $f_{3}$ and $g_{3}$ have been neglected which is because of the final state $e$ mass which is small and can be neglected. Using the current experimental values of the masses of initial and final baryons, branching fractions and lifetime of the decays from \cite{PDG}, we get the decay rates for the various decays discussed above. Since the data from experiment is accessible only for a few of these  decays where strangeness changes, $V_{us}$ can be calculated only for those cases. $\Xi^{-}  \rightarrow \Xi^{0} e^{-}\bar{\nu}_{e}$ has not been experimentally studied, however, the extremity on its branching ratio is fixed to a value $B(\Xi^{-}  \rightarrow \Xi^{0} e^{-}\bar{\nu}_{e}) < 2.59 \times 10^{-4} $  \cite{BESIII:2021emv}. The results for $V_{us}$ and $V_{ud}$ are presented in the Tables \ref{tab:t1} and \ref{tab:t2}. 

After using the form factors to calculate $V_{us}$ and $V_{ud}$, we go on to study the effect of vulnerability of $Q^2$. As the experimental data for different decays are available at different  $Q^2$, we have to find the correspondence of our results calculated at $Q^2=0$ GeV$^2$. The most established and accepted form of parametrization at momentum transfer $Q^{2} \le 1$ is the dipole form. The  vector and axial-vector form factors using this parametrization are expressed as
\bea
f_i^{B_iB_f}(Q^2)&=&\frac{f_i^{B_iB_f}(0)}{\left( 1+\frac{Q^2}{M_{f_i}^2}\right)^2}, \nonumber \\
g_i^{B_iB_f}(Q^2)&=&\frac{g_i^{B_iB_f}(0)}{\left( 1+\frac{Q^2}{M_{g_i}^2}\right)^2},
\label{dipole}\eea
where $f_i^{B_iB_f}(0)$ and $g_i^{B_iB_f}(0)$ are the coupling constants for vector and axial-vector currents at $Q^2=0$. The dipole masses  are represented by $M_{f_i}$ and $M_{g_i}$ respectively for the vector and axial-vector cases. 

\section{Inputs}\label{s3}
The first group of variables used in the numerical computation of the vector and axial-vector form factors are the symmetry breaking variables which are further established in connection with the probabilities for transitions/fluctuations of quarks in the quark sea of a specific baryon. The fitting of these variables has already been carried out in the context of evaluating the experimentally measured quantities describing the flavor and spin structure of the baryons \cite{nsweak} and we will be using the same variables in the present calculations.

The second set of input variables to be fixed are the masses of initial and final state baryons the values of which have been presented in Tables \ref{tab:t1} and \ref{tab:t2} for all the decays studied here. Further, to examine the variation of vector and axial-vector form factors for $Q^{2} \le 1$ GeV, $M_{V}/M_{f_{i}}$ and  ${M_{A}/M_{g_i}}$ values are required. 
We have summarized the inputs for the case where strangeness is conserved ($\Delta S=0$ decays) \cite{Mateu:2005wi} and for case where strangeness changes ($\Delta S=1$ decays) in Table \ref{inputs}.  For the case of {$V_{us}$ and {$V_{ud}$, the input of total decay rate, defined in Eq. (\ref{Decayrate}), can be used. Here we have used the values of the decay rate of $B_i \rightarrow B_f l \bar{\nu}$ available in PDG \cite{PDG}.

		   \section{Results and Discussion}\label{s4}
		\subsection{The CKM matrix elements, $V_{us}$ and $V_{ud}$}
		Making use of the variables presented in the previous section, in Table \ref{tab:t1} we furnish the results for the vector and axial-vector form factors corresponding to the decays where the strangeness changes along with their  CKM matrix elements. Similarly, in Table \ref{tab:t2}, the results for form factors and corresponding CKM matrix elements for decays where strangeness is conserved have been provided. We would like to add here that SU(3) symmetry breaking has been included while computing the results presented in Table \ref{tab:t1} and Table \ref{tab:t2}. These results are all  valid at $Q^{2} = 0$. We present the vector ($f_1$), induced tensor ($f_2$), axial-vector ($g_1$) and the induced pseudotensor ($g_2$) form factors. From Table \ref{tab:t1}, as illustrated for the vector form factor $f_1$, we have
		\be
		f_1 (\Lambda  \rightarrow pe^{-}\bar{\nu}_{e}) < f_1(\Sigma^{-} \rightarrow ne^{-}\bar{\nu}_{e})< f_1(\Xi^{-}  \rightarrow \Sigma^{0} e^{-}\bar{\nu}_{e})<f_1(\Xi^{0}  \rightarrow \Sigma^{+}e^{-}\bar{\nu}_{e})<f_1(\Xi^{-}  \rightarrow \Lambda e^{-}\bar{\nu}_{e}),
		\ee
		and 
		\be f_1 (\Xi^{-}  \rightarrow \Lambda e^{-}\bar{\nu}_{e})= -f_1 (\Lambda  \rightarrow pe^{-}\bar{\nu}_{e}) and f_1(\Sigma^{-} \rightarrow ne^{-}\bar{\nu}_{e}) =- f_1(\Xi^{0}  \rightarrow \Sigma^{+}e^{-}\bar{\nu}_{e}). \ee
		From Table \ref{tab:t2} we have
		\be
		f_1( \Xi^{-}  \rightarrow \Xi^{0} e^{-}\bar{\nu}_{e})<f_1(\Sigma^{-} \rightarrow \Lambda e^{-}\bar{\nu}_{e})=f_1(\Sigma^{+} \rightarrow \Lambda e^{-}\bar{\nu}_{e})<f_1(n \rightarrow pe^{-}\bar{\nu}_{e})<f_1(\Sigma^{-} \rightarrow \Sigma^{0} e^{-}\bar{\nu}_{e}),
		\ee
		and  \be f_1 (n \rightarrow pe^{-}\bar{\nu}_{e})= -f_1 (\Xi^{-}  \rightarrow \Xi^{0} e^{-}\bar{\nu}_{e}) and f_1(\Sigma^{-} \rightarrow \Lambda e^{-}\bar{\nu}_{e}) =f_1 (\Sigma^{+} \rightarrow \Lambda e^{-}\bar{\nu}_{e}). \ee 
		
		As illustrated for the form factor $f_2$ which arises due to the induced tensor current, we have from Table \ref{tab:t1}
		\be
		f_2 (\Lambda  \rightarrow pe^{-}\bar{\nu}_{e}) < f_2(\Xi^{-}  \rightarrow \Lambda e^{-}\bar{\nu}_{e}) <
		f_2(\Sigma^{-} \rightarrow ne^{-}\bar{\nu}_{e})< 
		f_2(\Xi^{-}  \rightarrow \Sigma^{0} e^{-}\bar{\nu}_{e})<
		f_2(\Xi^{0}  \rightarrow \Sigma^{+}e^{-}\bar{\nu}_{e}),
		\ee
		and from Table \ref{tab:t2} we have
		\be
		f_2(\Sigma^{-} \rightarrow \Sigma^{0} e^{-}\bar{\nu}_{e})<f_2( \Xi^{-}  \rightarrow \Xi^{0} e^{-}\bar{\nu}_{e})<f_2(\Sigma^{+} \rightarrow \Lambda e^{-}\bar{\nu}_{e})< f_2(n \rightarrow pe^{-}\bar{\nu}_{e})< f_2(\Sigma^{-} \rightarrow \Lambda e^{-}\bar{\nu}_{e}).
		\ee
		As emphasized earlier, for all the decays where strangeness changes as well as where strangeness is conserved (Tables \ref{tab:t1} and \ref{tab:t2}), the second class currents are quantitatively much smaller in magnitude in contrast to the first class currents. In addition, when the initial and final state baryons have the same isospin leading to small mass difference, the magnitude of $g_2$ is quite small. This is clear from the results of  $g_2$ presented in Table \ref{tab:t2} for the case of  $n \rightarrow pe^{-}\bar{\nu}_{e}$, $\Sigma^{-} \rightarrow \Sigma^{0} e^{-}\bar{\nu}_{e}$ and $\Xi^{-}  \rightarrow \Xi^{0} e^{-}\bar{\nu}_{e}$ decays. For the other cases, the value of $g_2$ increases with the increasing mass difference in initial and final state baryons.
		
		In the matter of axial-vector form factor $g_1$ we find from Table \ref{tab:t1} and \ref{tab:t2}
		\be
		g_1 (\Lambda  \rightarrow pe^{-}\bar{\nu}_{e}) < g_1(\Xi^{-}  \rightarrow \Lambda e^{-}\bar{\nu}_{e})< g_1(\Sigma^{-} \rightarrow ne^{-}\bar{\nu}_{e}) < g_1(\Xi^{-}  \rightarrow \Sigma^{0} e^{-}\bar{\nu}_{e}) < g_1(\Xi^{0}  \rightarrow \Sigma^{+}e^{-}\bar{\nu}_{e}),
		\ee
		\be
		g_1( \Xi^{-}  \rightarrow \Xi^{0} e^{-}\bar{\nu}_{e}) < g_1(\Sigma^{+} \rightarrow \Lambda e^{-}\bar{\nu}_{e})=g_1(\Sigma^{-} \rightarrow \Lambda e^{-}\bar{\nu}_{e})< g_1(\Sigma^{-} \rightarrow \Sigma^{0} e^{-}\bar{\nu}_{e}) <g_1(n \rightarrow pe^{-}\bar{\nu}_{e}).
		\ee
		Regarding induced pseudotensor (weak electricity) form factor $g_2$ we find from Table \ref{tab:t1} and \ref{tab:t2}
		\be
		g_2 (\Lambda  \rightarrow pe^{-}\bar{\nu}_{e}) < g_2(\Sigma^{-} \rightarrow ne^{-}\bar{\nu}_{e}) < g_2(\Xi^{-}  \rightarrow \Lambda e^{-}\bar{\nu}_{e})<  g_2(\Xi^{-}  \rightarrow \Sigma^{0} e^{-}\bar{\nu}_{e}) < g_2(\Xi^{0}  \rightarrow \Sigma^{+}e^{-}\bar{\nu}_{e}),
		\ee
		\be
		g_2(\Sigma^{-} \rightarrow \Lambda e^{-}\bar{\nu}_{e})< g_2(\Sigma^{+} \rightarrow \Lambda e^{-}\bar{\nu}_{e}) < g_2(\Sigma^{-} \rightarrow \Sigma^{0} e^{-}\bar{\nu}_{e}) < g_2( \Xi^{-}  \rightarrow \Xi^{0} e^{-}\bar{\nu}_{e}) <  g_2(n \rightarrow pe^{-}\bar{\nu}_{e}).
		\ee
		%%%%%%%%%%%%%%%%%%%%%%%%%%%%%%%%%%%%%%%%%%%%%%%%%%%%%%%%%%%%%%%%%%%%%%%%%%
		The decay $\Xi^{-}  \rightarrow \Sigma^{0} e^{-}\bar{\nu}_{e}$ gives
		$V_{us}$ closest to
		experimental value i.e. $0.2245 \pm 0.0008$.
		$\Sigma^{-} \rightarrow ne^{-}\bar{\nu}_{e} $ also gives $V_{us}$,
		which is within 1$\sigma$  of its experimental value.
		For the other three decays the values of $V_{us}$ obtained are far from
		the
		experimental value.
		
		The value of $V_{ud}$ given in Table \ref{tab:t2}, obtained from strangeness
		conserving decays is far from its
		experimental value i.e. $0.97370 \pm 0.00014$. The beta decay $n
		\rightarrow pe^{-}\bar{\nu}_{e}$ gives us the value that is particularly far from the experimental value. 
		%Also I did not see any pattern of $V_{ud}$ or $V_{us}$ dependence of
		%mass difference of initial and final
		%  state or any other parameter.
		%%%%%%%%%%%%%%%%%%%%%%%%%%%%%%%%%%%%%%%%%%%%%%%%%%%%%%%%%%%%%%%%%%%%%%%%
		
		In Table \ref{tab:t3} and \ref{tab:t4}, we present the decay constant ratio $\frac{f_2(0)}{f_1 (0)}$ for the decays where strangeness changes with $\Delta S=1$ and for the case when strangeness is conserved with  $\Delta S=0$ respectively. The experimental data is available only for the  $\Sigma^{-} \rightarrow ne^{-}\bar{\nu}_{e} $ and $\Xi^{0}  \rightarrow \Sigma^{+}e^{-}\bar{\nu}_{e}$ strangeness changing decays. Our results for $\Xi^{0}  \rightarrow \Sigma^{+}e^{-}\bar{\nu}_{e}$ are quite close to the data whereas $\frac{f_2(0)}{f_1 (0)}$ for the decay $\Sigma^{-} \rightarrow ne^{-}\bar{\nu}_{e} $ differ substantially. This difference comes from the fact that while predicting the data, the second class currents were assumed to be absent. Including the second class currents in future experiments would provide important implications for SU(3) symmetry breaking in this energy regime. 
		In case of the strangeness conserving decays, $\frac{f_2(0)}{f_1 (0)}$ for $\Sigma^{-} \rightarrow \Lambda e^{-}\bar{\nu}_{e}$ decay and 
		$\Sigma^{+} \rightarrow \Lambda e^{-}\bar{\nu}_{e}$ decay  cannot be defined as $f_1(0)=0$ for these cases. 
		%%%%%%%%%%%%%%%%%%%%%%%%%%%%%%%
		%%%%%%%%%%%%%%%%%%%%%%%%%
		%.......comparison with other models.......
		%%%%%%%%%%%%%%%%%%%%%%%%%%%
		%%%%%%%%%%%%%%%%%%%%%%%%%%%
		%%%%%%%%%%%%%%%%%%%%%%%%%%%%%%%%%%%%%%%%%%%%%%%%%%%%%%%%%%%%%%%%%%%%%%%%%%%%%%
		
		In Table \ref{tab:t5} and \ref{tab:t6}, we have presented the ratio of the axial-vector $g_1(0)$ and the vector form factor $f_1(0)$ for the decays where strangeness changes with $\Delta S=1$ and for the case when strangeness is conserved with  $\Delta S=0$ respectively. The ratio ${g_{1}(0) \over f_{1}(0)}$ basically gives the experimentally measured quantity ${g_{A} \over g_{V}}$ that has a direct relation with the quark spin polarization functions as well as the axial-vector couplings. These ratios are very important to completely understand the spin structure of the hadrons in general. The results are in good agreement for $\Sigma^{-} \rightarrow ne^{-}\bar{\nu}_{e} $, $\Xi^{-}  \rightarrow \Lambda e^{-}\bar{\nu}_{e}$, $\Lambda  \rightarrow pe^{-}\bar{\nu}_{e} $ and $\Xi^{0}  \rightarrow \Sigma^{+}e^{-}\bar{\nu}_{e}$  in the $\Delta S=1$ case and $n \rightarrow pe^{-}\bar{\nu}_{e}$ decay in the $\Delta S=0$ case. For the case of the strangeness conserving decay	$\Sigma^{-} \rightarrow \Lambda e^{-}\bar{\nu}_{e}$, the experimental data is available for ${f_{1} \over g_{1}}$ but since $f_1$ is $0$ in this case, we have presented the result for $g_{1}$ here. Similarly, in the case of $\Sigma^{+} \rightarrow \Lambda e^{-}\bar{\nu}_{e}$ the result of $g_{1}$ has been presented in the table. 
		%%%%%%%%%%%%%%%%%%%%%%%%%%%%%%
		%%%%%%%%%%%%%%%%%%%%%%%%%%%%%
		%%%%%%%%%%%%%%%%%%%%%%%%%%%
		%.......comparison with other models.......
		%%%%%%%%%%%%%%%%%%%%%%%%%%%%%%%%%%%%%%%
		%%%%%%%%%%%%%%%%%%%%%%%%%%%%%%%%%%%%%%%
		%%%%%%%%%%%%%%%%%%%%%%%%%%%%%%%%%%%%%%%%%%%
		This ratio is an important quantity in obtaining a consensus in the context of hadron dynamics particularly in the low energy regime where other degrees of freedom are required to describe the flavor and spin structure.
		%%%%%%%%%%%%%%%%%%%%%%%%%%%%%%%%%%%%%%%%%%%%%%%%%%%%%%%%%%%
		For the sake of completeness, in Table \ref{tab:t7} and \ref{tab:t8}, we have presented the  ratio ${g_{2}(0) \over g_{1}(0)}$ 
		for the decays where strangeness changes with $\Delta S=1$ and for the case when strangeness is conserved with  $\Delta S=0$ respectively. 
		%%%%%%%%%%%%%%%%%%%%%%%%%%%%%%%%%%%%%%%%%%%%%%%%%%%%%%%%%%%%%%%%%%%%%%%%

		%%%%%%%%%%%%%%%%%%%%%%%%%%%%%%%%%%%%ERROR in the table%%%%%%%%%%
		
		%&&&&&&&&&&&&&&&&&&&&&&&&&&&&&&&&&&&&&&&&&&&&&&&&&&&&&&&&&&&&&&&&&&&&&&&&&&&&&&&&&&&&&&&&&&&&&&&&&&&&&&&&&&&&&&
		
		%=================================================================================================================================
		\subsection{Vulnerability of $Q^2$ in vector and axial-vector form factors}
		%%%%%%%%%%%%%%%%%%%%%%%%%%%%%%%%%%%%%%%%%%%%%%%%%%
		We will now study the dependence of the vector ($f_1$) and induced tensor ($f_2$) form factors on $Q^2$ as presented in Figs. \ref{fig:f1s01} and \ref{fig:f2s01} respectively. The dependence of $f_1$ and $f_2$ has been presented for the decays where strangeness changes with $\Delta S=1$ and for the case when strangeness is conserved with  $\Delta S=0$ for finite $0 \leq Q^2 \leq 1$ GeV$^2$ using the dipole type of parametrization which is experimentally investigated through the pion electroproduction or quasielastic neutrino scattering.  Similarly, in Figs. \ref{fig:g1s01} and \ref{fig:g2s01}, we have exhibited the variation of $g_1$ and $g_2$  for the decays where strangeness changes with $\Delta S=1$ and for the case when strangeness is conserved with  $\Delta S=0$ for finite $0 \leq Q^2 \leq 1$. There is ample evidence of the quark sea dominance in the region where $Q^2$ is lower, whereas at higher  $Q^2$, the valence quarks are  prevalent. In general, the vulnerability of $Q^2$ is more intense and prominent at lower values of $Q^2$ as compared to high  $Q^2$ values.  In fact $f_1,\,f_2$, $g_1$ and $g_2$ approach 0 as $Q^2 \rightarrow 1$ GeV$^2$. This makes the underlying dynamics of the role of quark sea even more important. The form factors also directly depend on the quark as well as baryon masses. This is evident from the $Q^2=0$ GeV$^2$ values of the form factor. As the difference of the initial and final state baryon masses increases, the magnitude scales up. This is true for both the decays where strangeness changes with $\Delta S=1$ and when strangeness is conserved with  $\Delta S=0$. As  $Q^2$ increases, the form factor magnitudes start decreasing with the drop in values being more steep and abrupt for form factor with higher values at $Q^2=0$ GeV$^2$. 
		
		From the plots, one can easily discuss
		the variation and sensitivity to $Q^2$ for the vector $f_1(Q^2)$ and induced tensor $f_2(Q^2)$ form factors. From Fig. \ref{fig:f1s01} (a) for $\Delta S=1 $ decays, we observe that the value of $f_{1}$ for $\Xi^{-} \rightarrow \Lambda e^{-}\bar{\nu}_{e}$, $\Xi^{-} \rightarrow \Sigma^{0} e^{-}\bar{\nu}_{e}$ and $\Xi^{0} \rightarrow \Sigma^{+}e^{-} \bar{\nu}_{e}$  decreases with increasing $Q^{2}$ whereas $f_{1}$ increases with increasing $Q^{2}$ for $\Sigma^{-}\rightarrow n e^{-}\bar{\nu}_{e} $ and $\Lambda \rightarrow p e^{-} \bar{\nu}_{e}$. Similarly, for the $\Delta S=0 $ decays  in Fig. \ref{fig:f1s01} (b), the value of $f_{1}$ for 	$n \rightarrow pe^{-}\bar{\nu}_{e}$ and $\Sigma^{-} \rightarrow \Sigma^{0} e^{-}\bar{\nu}_{e}$ decreases with increase in $Q^{2}$ whereas the value of $f_{1}$ for $\Xi^{-} \rightarrow \Xi^{0}e^{-}\bar{\nu}_{e}$ increases with increase in $Q^{2}$. 
		In Fig. \ref{fig:f2s01} (a), for  the $\Delta S=1 $ decays, $\Sigma^{-} \rightarrow ne^{-}\bar{\nu}_{e} $ and  $\Lambda  \rightarrow pe^{-}\bar{\nu}_{e} $ $f_2$ increases with increasing $Q^{2}$ whereas for 	$\Xi^{-}  \rightarrow \Lambda e^{-}\bar{\nu}_{e}$, 	$\Xi^{-}  \rightarrow \Sigma^{0} e^{-}\bar{\nu}_{e}$ and 	$\Xi^{0}  \rightarrow \Sigma^{+}e^{-}\bar{\nu}_{e}$ it decreases with increasing $Q^{2}$. In Fig. \ref{fig:f2s01} (b), for  the $\Delta S=0 $ decays,  all the decays decrease with increasing $Q^2$, however, owing to almost same mass difference,} the curves corresponding $\Sigma^{+} \rightarrow \Lambda e^{-}\bar{\nu}_{e} $, $\Sigma^{-} \rightarrow \Lambda e^{-}\bar{\nu}_{e}$ and 
	$\Xi^{-}  \rightarrow \Xi^{0} e^{-}\bar{\nu}_{e}$ are very close and overlap. The vector form factors have not been measured  experimentally so it is difficult to compare them with data.

Further, in the event of axial-vector form factors $g_1(Q^2)$, $\Delta S=1 $ decays appearing in Fig. \ref{fig:g1s01},   the form factors in general with positive value at $Q^2=0$ GeV$^2$ fall as the $Q^2$ value increases, for example, see variations for $\Sigma^{-} \rightarrow ne^{-}\bar{\nu}_{e} $, 
$\Xi^{-}  \rightarrow \Lambda e^{-}\bar{\nu}_{e}$, 
$\Xi^{-}  \rightarrow \Sigma^{0} e^{-}\bar{\nu}_{e}$ and 
$\Xi^{0}  \rightarrow \Sigma^{+}e^{-}\bar{\nu}_{e}$. On the other hand, for the form factors with negative amplitude, as in $\Lambda  \rightarrow pe^{-}\bar{\nu}_{e} $, the amplitude  of $g_1$ increases as $Q^2$ increases. Since the amplitude  of $g_1$ at $Q^2=0$ GeV$^2$ for $\Sigma^{-} \rightarrow ne^{-}\bar{\nu}_{e} $ and   $\Xi^{-}  \rightarrow \Lambda e^{-}\bar{\nu}_{e}$ decay is small, the fall of the curve is smooth. 
On the other hand, for the decays when strangeness is conserved with  $\Delta S=0$, $g_{1}$ decreases with $Q^2$ as the amplitude at $Q^2=0$ GeV$^2$ is positive in all the cases. Further it is clear from Fig. \ref{fig:g2s01} that the induced pseudotensor form factors $g_2(Q^2)$ for the decays where strangeness changes with $\Delta S=1 $,  the amplitude of the form factors are minuscule for $\Sigma^{-} \rightarrow ne^{-}\bar{\nu}_{e} $ decay and $\Xi^{-}  \rightarrow \Lambda e^{-}\bar{\nu}_{e}$ decay leading to a negligible variation. For $\Lambda  \rightarrow pe^{-}\bar{\nu}_{e} $, there is an increasing trend for the form factors with increasing $Q^2$ whereas for $\Xi^{-}  \rightarrow \Sigma^{0} e^{-}\bar{\nu}_{e}$ and $\Xi^{0}  \rightarrow \Sigma^{+}e^{-}\bar{\nu}_{e}$ the trend is reversed. For the decays when strangeness is conserved with $\Delta S=0 $ decays, the form factors for $\Sigma^{+} \rightarrow \Lambda e^{-}\bar{\nu}_{e}$ and $\Sigma^{-} \rightarrow \Lambda e^{-}\bar{\nu}_{e} $ show an  increasing trend with $Q^2$ whereas the trend is slow for the other decays which is mainly because of the  small values they have at $Q^2=0$ GeV$^2$. 

Finally, for the $g_{1}(Q^2)/f_{1}(Q^2)$ presented in Fig. \ref{fig:g1f1s01} we find that since the ratio $g_{1}(0)/f_{1}(0)$ is same for $\Xi^{-} \rightarrow \Sigma^{0}e^{-}\bar{\nu}_{e}$, $\Xi^{0}  \rightarrow \Sigma^{+}e^{-}\bar{\nu}_{e}$ in the $\Delta S=1 $ decays and $n \rightarrow p e^{-}\bar{\nu}_{e}$ in  the $\Delta S=0 $ decays, the variation with $Q^2$ is also the same. In general, the overall variation of the ratio depends on the variation of $g_1$ and $f_1$ of that particular  decay. For $\Delta S=1 $ decays, the ratio increases with increasing $Q^2$ for $\Xi^{-}  \rightarrow \Sigma^{0} e^{-}\bar{\nu}_{e}$, $\Xi^{-}  \rightarrow \Lambda e^{-}\bar{\nu}_{e}$,   $\Lambda  \rightarrow pe^{-}\bar{\nu}_{e} $ and $\Xi^{0}  \rightarrow \Sigma^{+}e^{-}\bar{\nu}_{e}$. For $\Sigma^{-} \rightarrow ne^{-}\bar{\nu}_{e} $, it however decreases. When compared with other models, the sign of ratio $g_{1}(0)/f_{1}(0)$ for $\Sigma^{-} \rightarrow ne^{-}\bar{\nu}_{e}$ is negative  as opposed to the positive sign observed in neutron beta decay as is the case in Cabibbo model.
%%%%%%%\cite{Cabibbo1}.
For $\Delta S=0$ decays, the ratio can be discussed only for $n \rightarrow pe^{-}\bar{\nu}_{e}$, $\Sigma^{-} \rightarrow \Sigma^{0} e^{-}\bar{\nu}_{e}$ and $\Xi^{-}  \rightarrow \Xi^{0} e^{-}\bar{\nu}_{e}$ decays as $f_1$ is 0 for the other decays. In this case, for the first two decays $g_{1}/f_{1}$ increases with increasing $Q^2$ whereas for the third decay it decreases.

%%%%%%%%%%%%%%%%%%%%%%%%%%%%%%%%%%%%%%%%%%%%%%%%%%%%%%%%

\section{Summary and Outlook}\label{s5}
In the present work we have presented an analysis of vulnerability of $Q^2$ for the form factors $f_{i}$ and $g_{i}$ using dipole form of parametrizations. The results for this study are presented in Figs. \ref{fig:f1s01} - \ref{fig:g1f1s01}. We also calculated the CKM matrix elements $V_{us}$ and $V_{ud}$ using the latest experimental data of decay rate and the form factors calculated in the framework of Chiral Constituent Quark Model. The results are presented in the  Tables \ref{tab:t1} and \ref{tab:t2}. We find that the values of both $V_{ud}$ and $V_{us}$ obtained from Hyperon semileptonic decays contain large uncertainties and still can not compete with the more precise values obtained from other sources.
%\pagebreak
%*********************************************************************** 
\section*{ACKNOWLEDGMENTS}

H.D. would like to acknowledge the SERB, Department of Science and Technology, Government of India through the grant (Ref No.TAR/2021/000157) under TARE scheme for financial support. AG would like to thank Women in Science scheme, from Department of Science and Technology, Government of India for financial support through grant (WOS-A/SR/PM-106/2017) under . 

%\pagebreak

\begin{table}[h]
	%  \begin{centre}
	\begin{tabular}{|c|c|c|c||c|c|c|c|c|}
		\hline
		Decay&$M_i(GeV)$&$M_f(GeV)$&R($GeV$)&$f_1$&$f_2$&$g_1$&$g_2$&$V_{us}$\\
		\hline
		
		$\Sigma^{-} \rightarrow ne^{-}\bar{\nu}_{e} $ &1.197&0.939&$4.531 \times{10^{-18}}$& $-1.0$&1.813&0.314&0.017&$0.22416$\\
		\hline
		$\Xi^{-}  \rightarrow \Sigma^{0} e^{-}\bar{\nu}_{e}$&1.321 &1.192 & $3.498 \times{10^{-19}}$&0.707&2.029&0.898&0.310&$0.22452$\\
		\hline
		$\Xi^{-}  \rightarrow \Lambda e^{-}\bar{\nu}_{e}$ &1.321 &1.116&$2.264 \times {10}^{-18}$ &1.225&$-0.450$ &0.262&0.047&$0.23915$\\
		\hline
		$\Lambda  \rightarrow pe^{-}\bar{\nu}_{e} $&1.116&0.938&$2.083 \times {10}^{-18}$ &$-1.225$&$-1.037$&$-0.909$&$-0.170$&$0.21498$\\
		\hline
		
		$\Xi^{0}  \rightarrow \Sigma^{+}e^{-}\bar{\nu}_{e}$ & 1.315 & 1.189 & $5.726 \times 10^{-19}$  &1.0&2.854&1.27&0.446&$0.21526$ \\
		\hline
	\end{tabular}
	\caption{The decay constants $f_1 (Q^2=0)$, $f_2 (Q^2=0)$, $g_1 (Q^2=0)$, $g_2 (Q^2=0)$ and the CKM matrix element $V_{us}$ for the strangeness changing ($\Delta S=1$) decays.}
	\label{tab:t1}
	%  \end{centre}
\end{table}

\begin{table}[h]
	%  \begin{centre}
	\begin{tabular}{|c|c|c|c||c|c|c|c|c|}
		\hline
		Decay&$M_i(GeV)$&$M_f(GeV)$&R($GeV$)&$f_1(0)$&$f_2(0)$&$g_1(0)$&$g_2(0)$&$V_{ud}$\\
		\hline
		$n \rightarrow pe^{-}\bar{\nu}_{e}$ &0.939&0.938&$7.249 \times{10}^{-28}$&1.00&2.612&1.270 &$-0.004$&$1.30355$\\
		\hline
		$\Sigma^{-} \rightarrow \Sigma^{0} e^{-}\bar{\nu}_{e}$&1.197&1.192&$-$&1.414&1.033&0.676&$-0.010$&$-$\\
		\hline
		$\Sigma^{-} \rightarrow \Lambda e^{-}\bar{\nu}_{e} $&1.197&1.116&$2.553 \times{10^{-19}}$&0&2.265&0.646&$-0.152$&$0.91202$\\
		\hline
		$\Sigma^{+} \rightarrow \Lambda e^{-}\bar{\nu}_{e}$&1.189&1.116&$1.644 \times{10^{-19}}$&0&2.257&0.646&$-0.136$&$0.94797$\\
		\hline
		
		$\Xi^{-}  \rightarrow \Xi^{0} e^{-}\bar{\nu}_{e}$&1.322&1.315&$-$&$-1.00$&2.253&0.314&$-0.007$&$-$\\
		\hline
	\end{tabular}
	\caption{The decay constants $f_1 (Q^2=0)$, $f_2 (Q^2=0)$, $g_1 (Q^2=0)$, $g_2 (Q^2=0)$ and the CKM matrix element $V_{ud}$ for the strangeness conserving ($\Delta S=0$) decays.}
	\label{tab:t2}
	%  \end{centre}
\end{table}

\begin{table}[h]
	\begin{tabular}{|c|c|c|}
		\hline
		&$\Delta S=0$ decay & $\Delta S=1$ decay\\
		\hline
		$M_V/M_{g_{i}}$ & 0.84 $\pm$ 0.04 GeV&0.97GeV\\
		\hline
		$M_A/M_{f_{i}}$ & 1.08 $\pm$0.08 GeV&1.25 GeV\\
		\hline
	\end{tabular}
	\caption{Input variables to study the variation of $Q^2$ \cite{Gaillard:1984ny, Mateu:2005wi}.}
			\label{inputs}
		
		\end{table}

\begin{table}[h]
	%\begin{center}
	\begin{tabular}{|c|c|c|}
		\hline
		Decay&[exp]&$\frac{f_2(0)}{f_1 (0)}$ in  $\chi{CQM_{config}}$\\
		\hline
		
		$\Sigma^{-} \rightarrow ne^{-}\bar{\nu}_{e} $, & $-0.97 \pm 0.014$&$-1.813$\\
		\hline
		
		$\Xi^{-}  \rightarrow \Sigma^{0} e^{-}\bar{\nu}_{e}$&--&2.870\\
		\hline
		$\Xi^{-}  \rightarrow \Lambda e^{-}\bar{\nu}_{e}$&--&$-0.367$\\
		\hline
		$\Lambda  \rightarrow pe^{-}\bar{\nu}_{e} $&--&0.8465\\
		\hline
		$\Xi^{0}  \rightarrow \Sigma^{+}e^{-}\bar{\nu}_{e}$, & $2.0 \pm 0.09$& 2.854\\
		\hline
	\end{tabular}
	\caption{The decay constant ratio $\frac{f_2(0)}{f_1 (0)}$ for $\Delta S=1$ decays. }
	%\end{center}
	\label{tab:t3}
\end{table}

\begin{table}[h]
	\begin{tabular}{|c|c|c|}
		%\hline
		%$\Delta S &=& 0$ \\
		\hline
		Decay&[exp]&$\frac{f_2(0)}{f_1 (0)}$ in  $\chi{CQM_{config}}$\\
		\hline
		$n \rightarrow pe^{-}\bar{\nu}_{e}$&--&2.612\\
		\hline
		$\Sigma^{-} \rightarrow \Sigma^{0} e^{-}\bar{\nu}_{e}$&--&0.7306\\
		\hline
		$\Sigma^{-} \rightarrow \Lambda e^{-}\bar{\nu}_{e}$&--&--\\
		\hline
		$\Sigma^{+} \rightarrow \Lambda e^{-}\bar{\nu}_{e}$&--&--\\
		\hline
		$\Xi^{-}  \rightarrow \Xi^{0} e^{-}\bar{\nu}_{e}$&--&$-2.253$\\
		\hline
	\end{tabular}
	\caption{The decay constant ratio $\frac{f_2(0)}{f_1 (0)}$ for  $\Delta S=0$ decays. }
	%\end{center}
	\label{tab:t4}
\end{table}

\begin{table}[h]
	\begin{tabular}{|c|c|c|}
		%\hline
		%$\Delta S &=& 0$ \\
		\hline
		Decay&[exp]&$\frac{g_1(0)}{f_1 (0)}$ in  $\chi{CQM_{config}}$\\
		\hline
		$\Sigma^{-} \rightarrow ne^{-}\bar{\nu}_{e} $&$-0.340$ $\pm$ 0.017 &$-0.314$\\
		\hline
		$\Xi^{-}  \rightarrow \Sigma^{0} e^{-}\bar{\nu}_{e}$&--&1.270\\
		\hline
		$\Xi^{-}  \rightarrow \Lambda e^{-}\bar{\nu}_{e}$&$0.25$ $\pm$ 0.05 &0.214\\
		\hline
		$\Lambda  \rightarrow pe^{-}\bar{\nu}_{e} $&$0.718$ $\pm$ 0.015 & 0.742\\
		\hline
		$\Xi^{0}  \rightarrow \Sigma^{+}e^{-}\bar{\nu}_{e}$&1.22 $\pm$ 0.55 &1.27\\
		\hline
	\end{tabular}
	\caption{Ratio ${g_{A} \over g_{V}} = {g_{1}(0) \over f_{1}(0)}$ in our model and the corresponding latest experimental results for $\Delta S=1$ decays \cite{PDG}}
	\label{tab:t5}
\end{table}
%%%%%%%%%%%%%%%%%%%%%%%%%%%%%%%%%%%%%%%%%%%%%%%%%%%%%%%%%%%%%%%%%%%%%%%%%%%%%%

\begin{table}[h]
	\begin{center}
		\begin{tabular}{|c|c|c|}
			%\hline
			%$\Delta S &=& 0$ \\
			\hline
			Decay&[exp]&$\frac{g_1(0)}{f_1 (0)}$ in  $\chi{CQM_{config}}$\\
			\hline
			$n \rightarrow pe^{-}\bar{\nu}_{e}$& $1.2756$ $\pm$ 0.0013 &1.270\\
			\hline
			$\Sigma^{-} \rightarrow \Sigma^{0} e^{-}\bar{\nu}_{e}$&--&0.478\\
			\hline
			$\Sigma^{-} \rightarrow \Lambda e^{-}\bar{\nu}_{e}$& ${f_{1} \over g_{1}}= 0.01 \pm$ 0.10&$-$ \\
			\hline
			$\Sigma^{+} \rightarrow \Lambda e^{-}\bar{\nu}_{e}$&--&$-$\\
			\hline
			$\Xi^{-}  \rightarrow \Xi^{0} e^{-}\bar{\nu}_{e}$&--&$-0.314$\\
			\hline
		\end{tabular}
		\caption{Ratio ${g_{A} \over g_{V}} = {g_{1}(0) \over f_{1}(0)}$ in our model and the corresponding latest experimental results for $\Delta S=0$ decays \cite{PDG}}
		\label{tab:t6}
	\end{center}
\end{table}

\begin{table}[h]
	%\begin{center}
	\begin{tabular}{|c|c|c|}
		\hline
		Decay&[exp]&$\frac{g_2(0)}{g_1 (0)}$ in  $\chi{CQM_{config}}$\\
		\hline
		$\Sigma^{-} \rightarrow ne^{-}\bar{\nu}_{e} $&--&$0.0541$\\
		\hline
		$\Xi^{-}  \rightarrow \Sigma^{0} e^{-}\bar{\nu}_{e}$&--&0.3452\\
		\hline
		$\Xi^{-}  \rightarrow \Lambda e^{-}\bar{\nu}_{e}$&--&0.1794\\
		\hline
		$\Lambda  \rightarrow pe^{-}\bar{\nu}_{e} $&-- & 0.1870\\
		\hline
		
		$\Xi^{0}  \rightarrow \Sigma^{+}e^{-}\bar{\nu}_{e}$&--&0.3512\\
		\hline
	\end{tabular}
	\caption{Ratio $ {g_{2}(0) \over g_{1}(0)}$ in our model and the corresponding latest experimental results for $\Delta S=1$ decays \cite{PDG}}
	\label{tab:t7}
	%\end{center}
	
\end{table} 
%%%%%%%%%%%%%%%%%%%%%%%%%%%%%%%%%%%%%%%

\begin{table}[h]
	%\begin{center}
	\begin{tabular}{|c|c|c|}
		%\hline
		%$\Delta S &=& 0$ \\
		\hline
		Decay&[exp]&$\frac{g_2(0)}{g_1 (0)}$ in  $\chi{CQM_{config}}$\\
		\hline
		$n \rightarrow pe^{-}\bar{\nu}_{e}$&--&$-0.0031$\\
		\hline
		$\Sigma^{-} \rightarrow \Sigma^{0} e^{-}\bar{\nu}_{e}$&--&$-0.0148$\\
		\hline
		$\Sigma^{-} \rightarrow \Lambda e^{-}\bar{\nu}_{e}$&--&$-0.2353$ \\
		\hline
		$\Sigma^{+} \rightarrow \Lambda e^{-}\bar{\nu}_{e}$&--&$-0.2105$\\
		\hline
		$\Xi^{-}  \rightarrow \Xi^{0} e^{-}\bar{\nu}_{e}$&--&$-0.0223$\\
		\hline
	\end{tabular}
	\caption{Ratio ${g_{2}(0) \over g_{1}(0)}$ in our model and the corresponding latest experimental results for  $\Delta S=0$ decays \cite{PDG}}
	\label{tab:t8}
	%  \end{center}
\end{table}

\begin{figure}[h]
	\centering
	
	\includegraphics{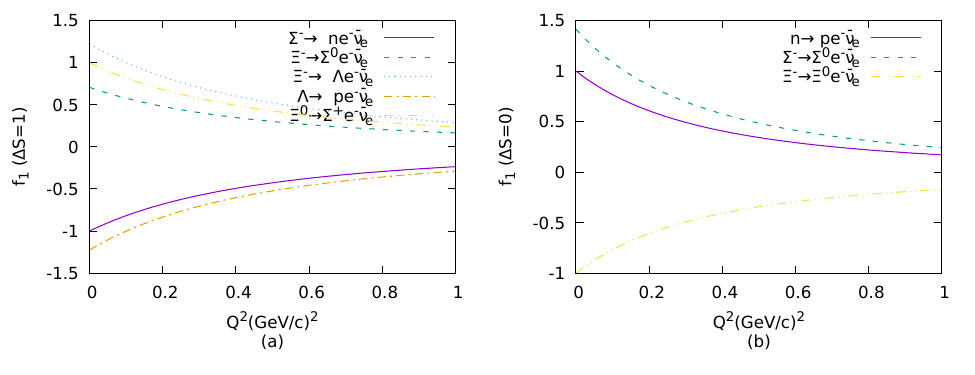} 
	
	\caption{Variation  of $f_{1}(Q^2)$ with $Q^{2}$ for $\Delta S=1$ and $\Delta S=0 $ decays.}
	\label{fig:f1s01}
\end{figure}

\begin{figure}[htb]
	
	\includegraphics{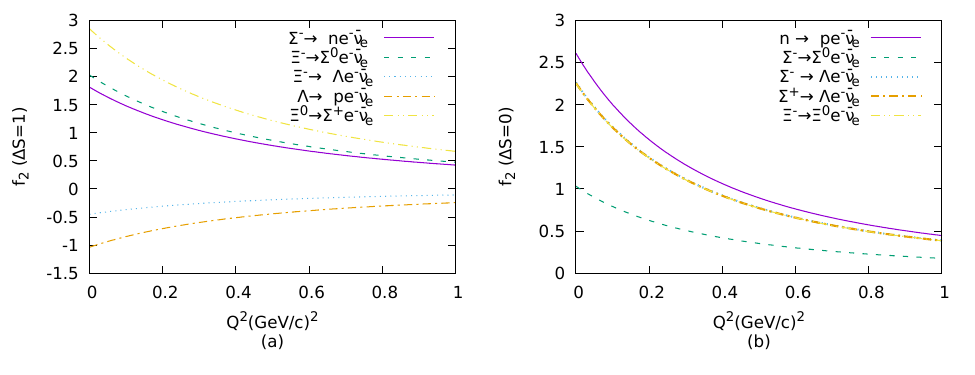} 
	
	\caption{Variation  of $f_{2}(Q^2)$ with $Q^{2}$ for $\Delta S=1 $ and $\Delta S=0 $ decays.}
	\label{fig:f2s01}
\end{figure} 
\begin{figure}[htb]	
	\includegraphics{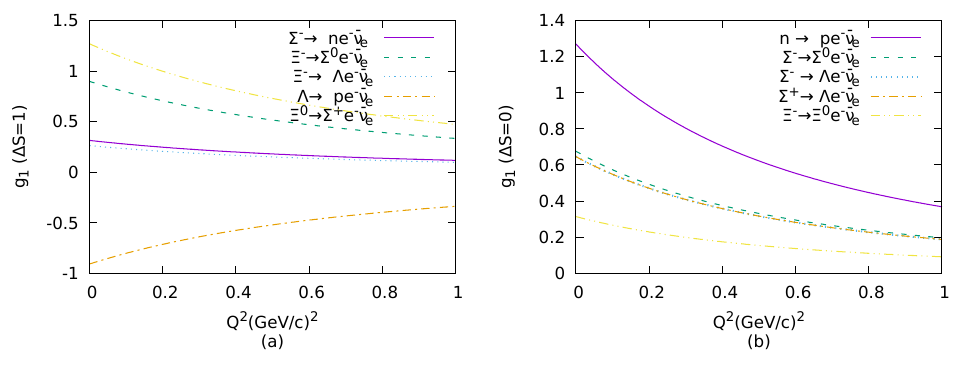} 
	
	\caption{Variation of $g_{1}(Q^2)$ with $Q^{2}$ for $\Delta S=1 $ and  $\Delta S=0 $ decays.}
	\label{fig:g1s01}
\end{figure}
%%%%%%%%%%%%%%%%%%%%%%%%%%%%%%%%%%%%%%%%%%%%%%%%%%%%%%%%%%%%%%%
\begin{figure}[htb]	
	\includegraphics{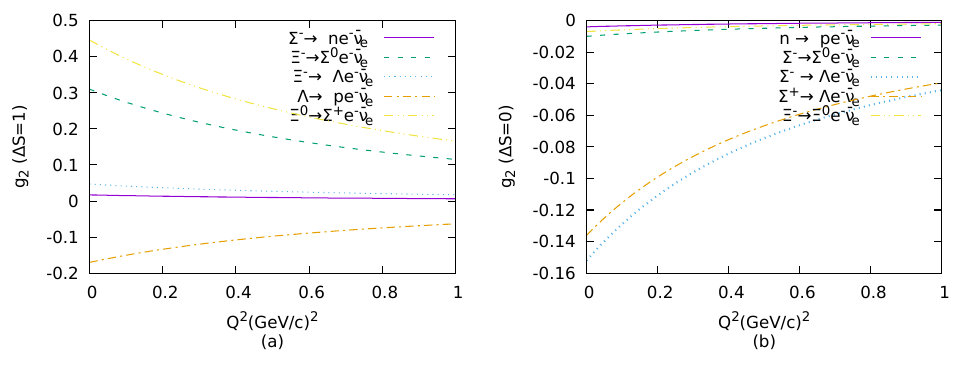} 
	\caption{Variation of $g_{2}(Q^2)$ with $Q^{2}$ for $\Delta S=1 $ and $\Delta S=0 $ decays.}
	\label{fig:g2s01}
\end{figure}

\begin{figure}[htb]
	\includegraphics{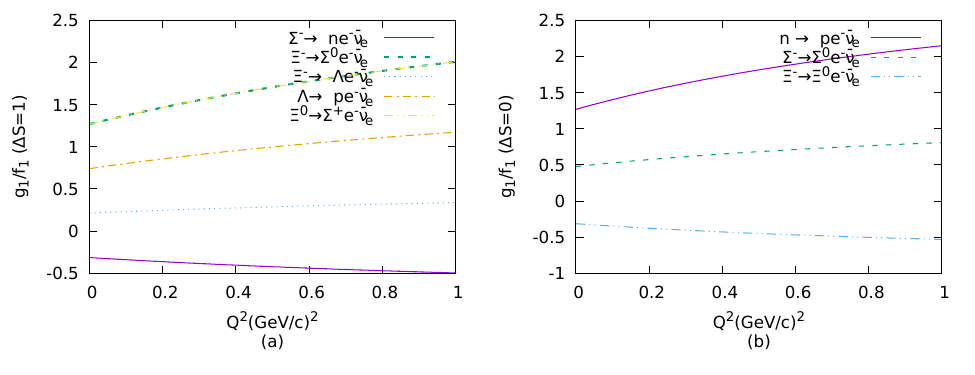} 
	
	\caption{Variation of $g_{1}(Q^2) \over f_{1}(Q^2)$ with $Q^{2}$ for $\Delta S=1$ and for $\Delta S=0 $ decays.}
	\label{fig:g1f1s01}
\end{figure}

\end{document}